\documentclass{article}

\usepackage[english]{babel}
\usepackage[toc, acronym, shortcuts]{glossaries}

\usepackage[letterpaper,top=2cm,bottom=2cm,left=3cm,right=3cm,marginparwidth=1.75cm]{geometry}

\usepackage{amsmath}
\usepackage{graphicx}
\usepackage{float}

\usepackage{subcaption}
\usepackage{csquotes}
\usepackage{url}
\usepackage[backend=bibtex]{biblatex}
\title{HTTPA: HTTPS Attestable Protocol}
\author{
Gordon King $^*$ \\
gordon.king@intel.com\\
Intel Corporation
\and
Hans Wang $^*$\\
hans.wang@intel.com\\
Intel Corporation
}

\date{}

\begin{document}
\maketitle

\def\thefootnote{*}\footnotetext{These authors contributed equally to this work.}

\begin{abstract}

Hypertext Transfer Protocol Secure (HTTPS) protocol has become an integral part of modern Internet technology. 
Currently, it is the primary protocol for commercialized web applications. 
It can provide a fast, secure connection with a certain level of privacy and integrity, and it has become a basic assumption on most web services on the Internet. 
However, HTTPS alone cannot provide security assurances on request data in computing, so the computing environment remains uncertain of risks and vulnerabilities. 
A hardware-based trusted execution environment (TEE) such as Intel\textsuperscript{\textregistered} Software Guard Extension (Intel\textsuperscript{\textregistered} SGX) or Intel\textsuperscript{\textregistered} Trust Domain Extensions (Intel\textsuperscript{\textregistered} TDX) provides in-memory encryption to help protect runtime computation to reduce the risk of illegal leaking or modifying private information. 
(Note that we use SGX as an example for illustration in the following texts.)
The central concept of SGX enables computation inside an enclave, a protected environment that encrypts the codes and data pertaining to a security-sensitive computation. 
In addition, SGX provides security assurances via remote attestation to the web client to verify, including TCB identity, vendor identity, and verification identity.
Here, we propose an HTTP protocol extension, called HTTPS Attestable (HTTPA), by including a remote attestation process onto the HTTPS protocol to address the privacy and security concerns on the web and the access of trust over the Internet. 
With HTTPA, we can provide security assurances for verification to establish trustworthiness with web services and ensure the integrity of request handling for web users. 
We expect that remote attestation will become a new trend adopted to reduce the security risks of web services.
We propose the HTTPA protocol to unify the web attestation and accessing Internet services in a standard and efficient way. 
\end{abstract}

\section{Introduction}
Privacy is deeply rooted in human rights as principles and protected by law. With recent hacks and breaches, more online consumers are aware of their private data being at risk. In fact, they have no effective control over it, leading to cybersecurity anxiety which is a new norm for our time. Obviously, the demands for cybersecurity and data privacy are rising. There are many interrelated efforts to protect sensitive data at rest, in transit, and in computing. 
Many of them have been applied to cloud, web services, and online businesses. 
Hypertext Transfer Protocol Secure (HTTPS)~\cite{uptls, https} is widely used to secure request data in motion, but the user data may be at risk i.e. data breach if the processing code is not fully isolated from everything else including the operating system on the host machine. A hardware-based TEE such as Intel\textsuperscript{\textregistered} SGX enclave is specifically designed for this concern. 
A remote system or users can get evidence through attestation to verify a trusted computing base (TCB) identity and state of TEE. Note that TCB includes hardware, firmware, and software to enforce security policy while TEE provides an application with secure isolation in run time on TCB. 
Most of the existing TEE solutions for protecting web services are very narrowly fit for addressing specific domain problems~\cite{selvi2014bypassing}. We propose a general solution to standardize attestation over HTTPS and establish multiple trusted connections to protect and manage requested data for selected HTTP~\cite{httpmr} domains. Also, our solution leverages the current HTTPS protocol, so it does not introduce much complexity like other approaches~\cite{amann2017mission}. This paper first discusses threat modeling. Then, we propose our protocol construction for attestation over HTTPS, which is called HTTPS Attestable (HTTPA). We hope this protocol can be considered a new standard in response to current Web security concerns. Lastly, we suggest that there are two operation modes to discover, including one-way HTTPA and mutual HTTPA (mHTTPA).

\section{Threat Modeling}
HTTPA is designed to provide assurances via remote attestation~\cite{attestguide} and confidential computing between a client and a server under the use case of the World Wide Web (WWW) over the Internet, so the end-point user can verify the assurances to build trust. 
For one-way (or unilateral) HTTPA, we assume the client is trusted and the server is not trusted, so the client wants to attest the server to establish trust. 
The client can verify those assurances provided by the server to decide whether they want to trust to run the computing workloads on the non-trusted server or not. However, HTTPA does not provide guarantees to make the server trustful. 
HTTPA involves two parts: communication and computation. 
Regarding communication security, HTTPA inherits all the assumptions of HTTPS for secure communication, including using TLS and verifying the host identity via a certificate. 
Regarding computation security, HTTPA protocol requires providing further assurance status of remote attestation for the computing workloads to run inside the secure enclave, so the client can run the workloads in encrypted memory with the proved enclave and proved TCB. 
As such, the attack surface of the computing is the secure enclave itself, and everything outside the secure enclave is not trusted. 
We assume that attackers on the server end have privileged access to the system with the ability to read, write, delete, replay and tamper with the workloads on the unencrypted memory. 
On the other hand, we assume software running inside the enclave is trusted if its certificate passes the client user's verification. 
Furthermore, HTTPA protocol requires that software vendor identity, TCB, and quote verification provider's identity are verifiable, so the protocol allows the client user to only trust what it agrees to trust by making its own allowed list and denied list. 
Also, the HTTPA provides an assurance to confirm the client's workloads to run inside the expected enclave with expected verified software.
As such, users can confirm their workloads are running inside the expected secure enclave in the remote and have the right to reject the computation when the verification result does not meet their security requirements. 
Running inside the secure enclave can significantly reduce the risks of user codes or data being read or modified by the attack from outside the enclave. 
Therefore, HTTPA can further reduce the attack surface of HTTPS from the whole host system to the secure enclave. 
Lastly, HTTPA provides freedom for users to determine whether they agree with the results of the assurances or not before proceeding to the computation and thus further reducing cyber-security risks. 

\section{Problem Statement}
Currently, many software services behind a website are still vulnerable due to unsafe memory computation. 
Lots of in-network processing even worsen the situation by enlarging the attack surface.
There is no sufficient assurance for workload computation such that most web services remain lacking in trust.
We argue that the current HTTPS is not sufficient at all to build fundamental trust for users in the modern cloud computing infrastructure.
The current situation of the Internet can cause people's private information and digital assets at great risk that users fully lose control over their own data.
Following the end-to-end principle, we propose an attestation-based assurance protocol at the application layer of the OSI model (or L7) to rescue.
Although we cannot guarantee fully secure computation, we can provide assurances to greatly reduce risks and regain control of data for Internet users where trust is built by verifying assurances.

\section{HTTPS Attestable (HTTPA) Protocol}
In this section, we first describe the standard HTTPS protocol on which we build our solution to establish an initial secure channel between the web server and the client. Then, we present the HTTPA protocol by adding a new HTTP method~\cite{httpsc} of attestation to establish multiple trusted channels. The new method unifies attestation~\cite{remoteattestsrv, dcapattest} and web access operations together for web applications in a general way.

\subsection{Standard HTTP over TLS}
The primary purpose of TLS~\cite{tls} is to protect web application data from unauthorized disclosure and modification when it is transmitted between the client and the server. The HTTPS/TLS security model uses a ``certificate'' to ensure authenticity. The certificate is cryptographically ``signed'' by a trusted certificate authority (CA)~\cite{cmp}. It is the current standard practice for communication over the Internet, 

\begin{figure}
\centering
\includegraphics[width=4in]{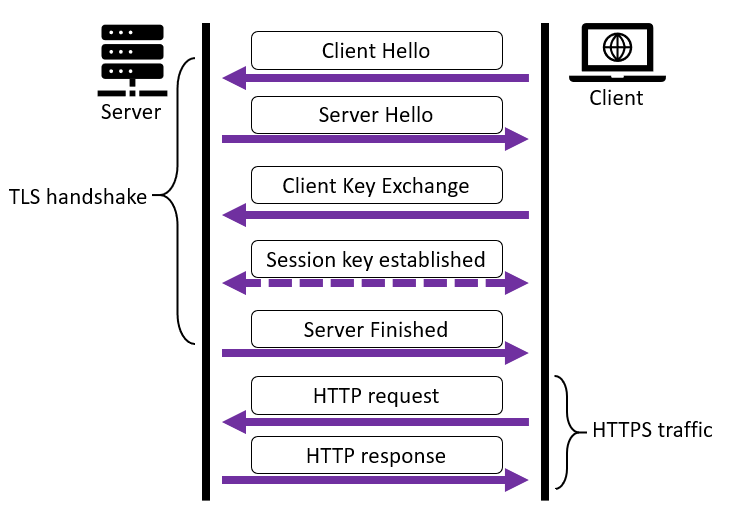}
\caption{An overview of standard HTTPS (HTTP over TLS) handshake process.}
\label{fig:regular_https_hs}
\end{figure}

Figure.\ref{fig:regular_https_hs} shows the HTTPS handshake mechanism. In HTTPS, the communication protocol is encrypted using TLS. It provides authentication of the server to the client as well as integrity guarantees, forward secrecy, and replay prevention. but it does not offer any of these security benefits to data that is in compute and at rest. Therefore, the request handling codes running in an untrusted environment which may expose sensitive data to attackers are not trusted. 
The client user cannot have enough confidence about their private information without being disclosed while the monitoring/notification processes are also running in an untrusted environment, as shown in Figure.\ref{fig:untrusted_service_access}.



Figure.\ref{fig:untrusted_service_access} shows the possible attacks if only using HTTPS to protect the request data sent by remote web users or systems. They can validate a certificate that represents the specific domain. To authorize a runtime environment is extremely difficult if not well protected by a hardware-based TEE because it is constantly changing. Therefore, the session keys, private key, and cleartext are possible to be revealed if they are handled in an untrusted environment.

\subsection{Attestation over HTTPS}
\begin{figure*}[htp]
  \begin{subfigure}[t]{0.48\textwidth}
    \includegraphics[width=\textwidth]{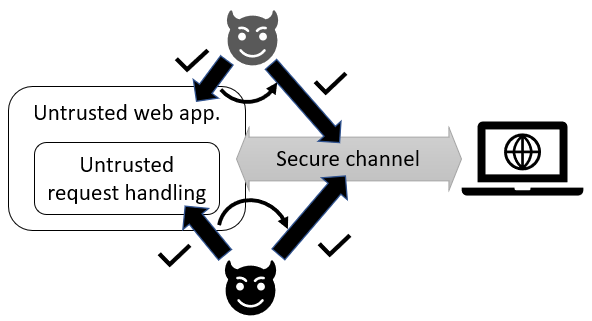}
    \caption{The current security status on the Internet by using HTTPS, which provides untrusted but secure service access. The attacker can perform a privileged escalation attack on the server to hack the session. Once the attack obtains the secret session, the attacker can compromise the secure channel between the server and the client by using the key to decrypt packets.  }
    \label{fig:untrusted_service_access}
  \end{subfigure}
  \hfill
  \begin{subfigure}[t]{0.48\textwidth}
    \includegraphics[width=\textwidth]{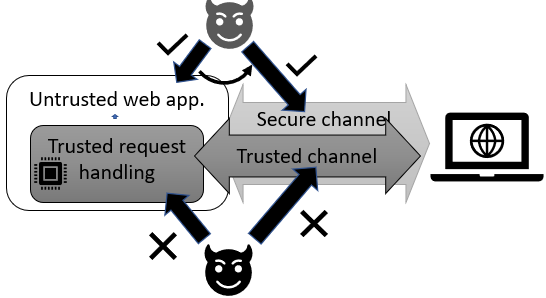}
    \caption{The possible future security status on the Internet by using the proposed HTTPA, which provides trusted and secure service access. The attacker cannot easily hack the session keys inside the secure enclave even though it has privileged access to the server. Therefore, it is more difficult for the attacker to compromise the privacy and integrity of the client's data on the server. }
    \label{fig:trusted_service_access}
  \end{subfigure}
  \hfill
  \caption{The proposed HTTPA protocol provides HTTP attestation methods to establish a trusted channel on top of the secure channel of HTTPS. Therefore, the HTTPA protocol significantly reduces the risks of exposing sensitive data than HTTPS does. }
\end{figure*}

The Intel\textsuperscript{\textregistered} SGX technology~\cite{intelsgx} is designed to provide hardware-based TEE to reduce the TCB and protect private data in computing. Smaller TCB implies reducing more security risks because it reduces the exposure to various attacks and the surface of vulnerability. Also, it can use an attestation service to establish a trusted channel. 
With hardware-based TEE, an attestation mechanism can help rigorously verify the TCB integrity, confidentiality, and identity. 
We propose to use remote attestation as the core interface for web users or web services to establish trust as a secure trusted channel to provision secrets or sensitive information. 
To achieve this goal, we add a new set of HTTP methods, including \emph{HTTP preflight~\cite{preflight} request/response, HTTP attest request/response, HTTP trusted session request/response}, to realize remote attestation which can allow web users and the web services for a trusted connection directly with the code running inside the hardware-based TEE, as shown in Figure.\ref{fig:trusted_service_access}

\begin{figure}
\centering
\includegraphics[width=\textwidth]{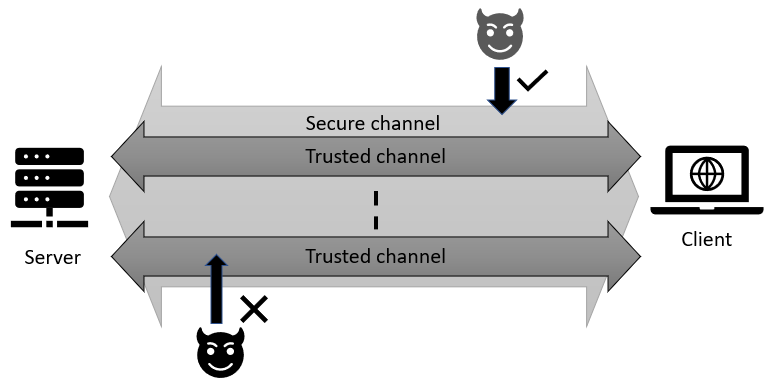}
    \caption{The proposed HTTPA supports establishing multiple trusted channels on top of the same secure channel as a scalable solution to the current Internet use case.}
    \label{fig:secure_trusted_channel}
\end{figure}
  

The TLS protocol supports the modification of the existing secure channel with the new server certificate and ciphering parameter; there are some efforts on exploring the way to weave the attestation process into TLS layer~\cite{knauth2019integrating}~\cite{bhardwaj2018spx}; however, our method does not replace existing secure channel with the trusted channel. Furthermore, we can allow the creation of multiple trusted channels inside a single TLS-secure channel, as shown in Figure.\ref{fig:secure_trusted_channel}.

\begin{figure}
\centering
\includegraphics[width=4in]{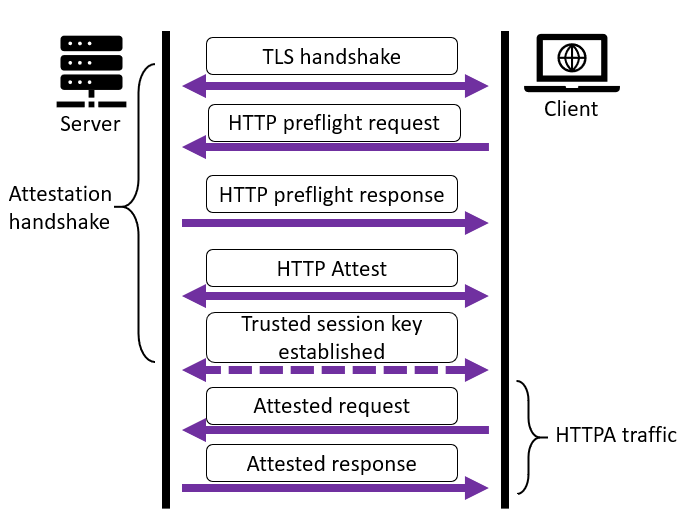}
\caption{An overview of the proposed HTTPA protocol. The HTTPA protocol consists of three sets of HTTP methods, including \emph{HTTP preflight request/response}, \emph{HTTP attest request/response}, and \emph{HTTP trusted session request/response}. }
\label{fig:httpa_hs_overview}
\end{figure}
\begin{figure}
\centering
\includegraphics[width=3in]{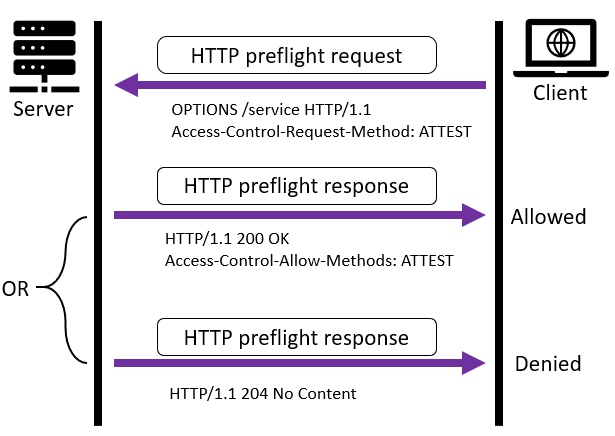}
\caption{ Following Figure.\ref{fig:httpa_hs_overview}, we specifically show how \emph{HTTP preflight request/response} works in details. First, the client sends out \emph{HTTP preflight request} to confirm whether the server has attestation capability. If the server does not have attestation capability, it responds with no content. Then the communication between the server and the client stops because the server does not have the capability to provide trusted assurances. If the server has attestation capability, it responds with OK to allow for the attestation method to proceed.}
\label{fig:httpa_preflight_hs}
\end{figure}

Figure.\ref{fig:httpa_hs_overview} shows an overview of proposed HTTPA handshake process. 
The HTTPA handshake process consists of three stages: HTTP preflight request/response, HTTP attestation request/response, and HTTP trusted session request/response.
HTTP preflight request/response will check whether the platform is attestation available, as shown in Figure.\ref{fig:httpa_preflight_hs}, has nothing to do with security, and they have no bearing on a web application. 
Rather, the preflight mechanism benefits servers that were developed without an awareness of HTTPA, and it functions as a sanity check between the client and the server that they are both HTTPA-aware. the following attestation request is not a simple request, for that, having preflight requests is kind of a ``protection by awareness''.
HTTP attestation request/response can provide a quote and quote verification features. 
HTTP trusted session request/response can establish a trusted session to protect HTTPA traffics in which only the verified TCB codes can see the request data. 
A trusted session that creates a trusted channel begins with the session key generated and ends with destroying it. 
The trusted session can temporarily store information related to the activities of HTTPA while connected.
We describe our HTTPA protocol in terms of one-way HTTPA and mutual HTTPA (mHTTPA) in the following sections.

\subsection{One-way HTTPA}

\begin{figure}
  \centering
  \includegraphics[width=4in]{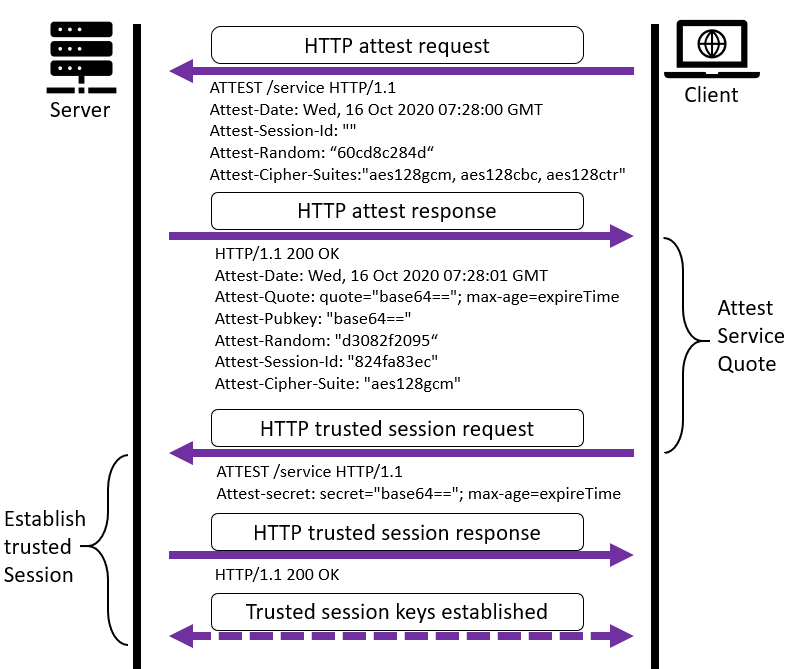}
  \caption{Following}
  \label{fig:httpa_attest_1w_hs}
\end{figure}

One-way HTTPA protocol is built on top of standard HTTPS. 
In this case, only the client validates the server to ensure that it receives data from the expected TCB and its hosting server. Our methodology is to perform the HTTP attestation handshake over the established secure channel of HTTPS. 
Our one-way HTTPA is described as follows. 
The client and server establish a secure channel if the preflight request for each domain is successful. 
Preflight request checks if the attestation protocol is accepted by the server for using the ``ATTEST'' method and headers. 
It is an ``OPTIONS request'', using one HTTP request header: Access-Control-Request-Method. 
Following the preflight request/response, a new set of HTTP methods is proposed to handle attestation message exchange, as shown in Figure.\ref{fig:httpa_attest_1w_hs}. 
The set of HTTP methods includes \emph{HTTP attest request}, \emph{HTTP attest response}, \emph{HTTP trusted session request}, \emph{HTTP trusted session response}

First, \emph{HTTP attest request} is generated by the client, using three new HTTP request headers as follows: 
\begin{enumerate}
\item Attest-Date \hfill \\
This item contains the date and time at which the attestation quote material was generated on the client side. This is optional for the one-way HTTPA protocol. 

\item Attest-session-ID \hfill \\
A unique string identifies a session, Here it is empty/null. If we had previously connected to a TEE a few seconds ago, we could potentially resume a session and avoid a full attestation handshake to happen again.

\item Attest-Random \hfill \\
This is a set of random bytes generated by the client for key derivation.

\item Attest-Cipher-Suites \hfill \\
This is a list of all of the encryption algorithms that the client is willing to support.
\end{enumerate}

\begin{figure}
\centering
\includegraphics[width=4in]{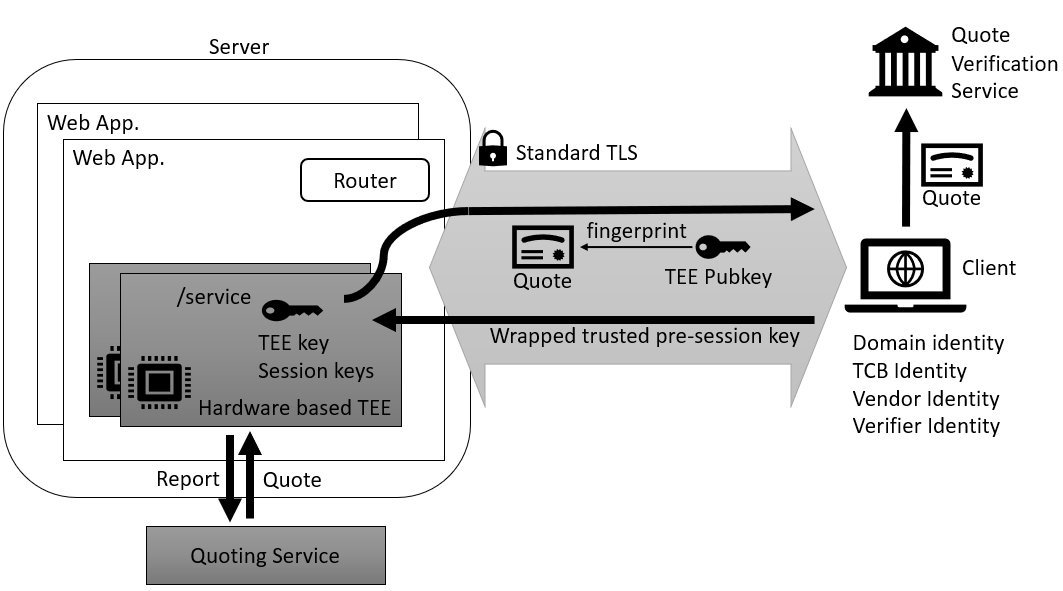}
\caption{The summary of the primary steps of HTTPA, including quote generation, third-party quote verification, obtaining TEE's public key, and the pre-session secret to establishing the trusted channel between the server and the client.}
\label{fig:httpa_oneway_attest}
\end{figure}

Second, the server will provide the received random bytes to the TEE once it got created and ready to accept requests.
The following is the \emph{HTTP attest response}, including the headers.
\begin{enumerate}
\item Attest-Date \hfill \\
The date and time at which the attestation quote material was generated on the server side.

\item Attest-Quote \hfill \\
This item contains a quote that was generated by a TCB hosting web service that will handle HTTPA requests to the domain in question. the max-age indicates how long the results of ATTEST request can be cached.

\item Attest-Pubkey \hfill \\
This item contains a public key that was generated inside the TEE for exchanging secret keys. Its paired private key should never leave its TEE. In addition, it needs to be bound to the identity of its TCB, so the fingerprint of the public key should be carried by its TCB quote as payload, see Figure.\ref{fig:httpa_oneway_attest}

\item Attest-Random \hfill \\
This item contains random bytes generated by the server. This will be used later. 

\item Attest-Session-Id \hfill \\
The session id is generated by a web service running inside the TEE of the server.

\item Attest-Cipher-Suite \hfill \\
This is an encryption algorithm picked by the server for trusted channel encryption.
\end{enumerate}


Third, the client uses the received public key of TEE to wrap a secret which is the pre-session. The wrapped pre-session secret is later used for the key derivation.
The web application should send the wrapped pre-session secret directly into the web service TEE because the secret can only be unwrapped by its paired private key inside the TEE of the web service.
\emph{HTTP trusted session request} includes the following headers:

\begin{enumerate}
    \item Attest-Secret \hfill \\
    Contains a pre-session secret wrapped by the server-side TEE public key.
\end{enumerate}

In the last step, \emph{HTTP trusted session response} confirms the pre-session secret has been received by the server.

Thus far, both server and client sides (and only those sides) have a pre-session secret. 
Each party can calculate the ``trusted session keys'', which are derived from the pre-session secret and the random bytes of both parties.
To create the trusted session keys, including MAC keys, encryption keys, and the IV for cipher block initialization, we need to use PRF, the ``Pseudo-Random Function'', to create a ``key block'' where we pull data from:
\begin{verbatim}
key_block = PRF(pre_session_secret, 
                    "trusted session keys", 
                    ClientAttest.random + ServerAttest.random)
\end{verbatim}
The pre-session secret is the secret we sent earlier, which is simply an array of bytes. 
We then concatenate the random bytes which are sent via HTTP attest request/response from both server and client sides.
We use PRF to combine the secrets with the concatenated random bytes from both server/client sides. 
The bytes from the ``key block'' are used to populate the following:

\begin{verbatim}
client_write_MAC_secret[size]
server_write_MAC_secret[size]
client_write_key[key_material_length]
server_write_key[key_material_length]
client_write_IV[IV_size]
server_write_IV[IV_size]
\end{verbatim}

The use of Initialization Vectors (IVs) depends on which cipher suite is selected at the very beginning, and we need two Message Authentication Code (MAC) keys for each side. 
In addition, both sides need the encrypt keys and use those trusted session keys to make sure that the encrypted request data has not been tampered with and can only be seen by the request handlers running in the attested TCB.

The client needs to verify the server quote to establish the trustworthiness of the platform, including verifying the integrity of the codes running inside the TCB of web services. 
Usually, there are authorities that offer such services, the client can request a trusted attestation service to verify the server quote. 
A verifier quote may be returned to the client along with the verification result; thus, the client side can collect several identities to make a decision, the router, shown in Figure.\ref{fig:httpa_oneway_attest}, could be placed in the un-trusted part of a web application to route service requests corresponding to HTTP header, the web applications is expected to create multiple isolated TCB to confidentially handle a different kind of requests.

\begin{enumerate}
\item Domain identity \hfill \\
It is the identity issued by a Certificate Authority (CA), and it is used in the TLS protocol

\item TCB identity \hfill \\
It is the identity of web service TCB, measured by hardware and it is embedded in a quote. 

\item Vendor identity \hfill \\
It is a signing identity of web service provided by a CA, which signs the TCB prior to distribution

\item Verifier identity \hfill \\
The identity is received from an attestation service along with the attesting result.
\end{enumerate}

All those identities need to be validated by verifying their certificate chain. The user can selectively add any combinations of them to the allowed list or denied list. 

\subsection{Mutual HTTPA (mHTTPA)}
In addition to the one-way HTTPA protocol, HTTPA includes another direction, so both client and server attest to each other to ensure that both parties involved in the communication are trusted. Both parties share their quotes on which the verification and validation are performed.

\begin{figure}[H]
\centering
\includegraphics[width=4in]{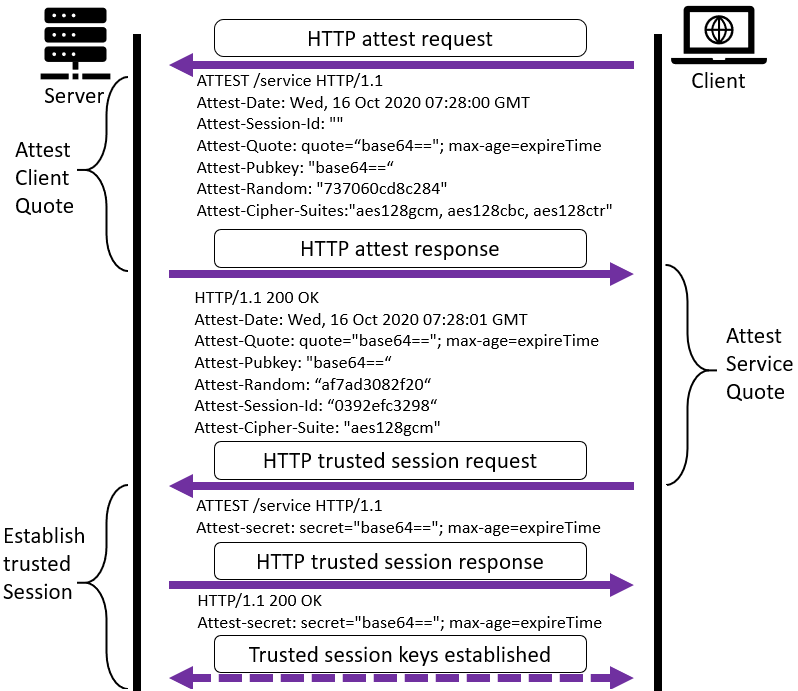}
\caption{Comparing with Figure.\ref{fig:httpa_attest_1w_hs}, the major difference of mHTTPA from one-way HTTPA is the step of \emph{HTTP attest request}. After confirming the server's capability to provide attestation described in Figure.\ref{fig:httpa_preflight_hs}, the client generates its own quote for the server to verify. 
Then the client sends \emph{HTTP attest request} to the server, which includes the client's generated quotes, TEE public key, random bytes, cipher suites, and meta information. 
Then the server verifies the quote of the client. 
If the client's quote is not verified successfully, the communication stops. 
If the quote is verified successfully, the server generates its quote. Then the server sends \emph{HTTP attest response} to respond to the client with the server's quote, TEE public key, random bytes, chosen cipher algorithm, and the meta information. 
After this part, the remaining steps follow the same as described in Figure.\ref{fig:httpa_attest_1w_hs} for each party.}
\label{fig:httpa_attest_2w_hs}
\end{figure}

Figure.\ref{fig:httpa_attest_2w_hs} shows the handshake process of mHTTPA protocol. Compared to one-way HTTPA, see Figure.\ref{fig:httpa_attest_1w_hs}, the difference is that client will send its quote and its TEE public key to the server in the first request to allow the server for verifying the quote, and then send a service quote back as a response if the client quote got attested.

There is a slight change to the key block generation as well, for now, both sides need to include two pre-session secrets for deriving session keys in their own TEEs respectively.
\begin{verbatim}
key_block = PRF(server_pre_session_secret, client_pre_session_secret,
                    "trusted mutual session keys", 
                    ClientAttest.random + ServerAttest.random)
\end{verbatim}

\section{Summary}
This paper presents the HTTPA as a new protocol to unify the attestation process and HTTPS. 
With HTTPA, we can establish a trusted channel over HTTPS for web service access in a standard and effective way. 
We demonstrate the proposed HTTPA protocol where attestation is used to provide assurances for verification, and we show how a web service is properly instantiated on a trusted platform. 
The remote web user or system can then gain confidence that only such intended request handlers are running on the trusted hardware. 
Using HTTPA, we believe that we can reduce security risks by verifying those assurances to determine the acceptance or rejection. 

\printbibliography

@article{selvi2014bypassing,
  title={Bypassing HTTP strict transport security},
  author={Selvi, Jose},
  journal={Black Hat Europe},
  volume={54},
  year={2014},
}

@inproceedings{amann2017mission,
  title={Mission accomplished? HTTPS security after DigiNotar},
  author={Amann, Johanna and Gasser, Oliver and Scheitle, Quirin and Brent, Lexi and Carle, Georg and Holz, Ralph},
  booktitle={Proceedings of the 2017 Internet Measurement Conference},
  pages={325-340},
  year={2017},
}

@misc{knauth2019integrating,
      title={Integrating Remote Attestation with Transport Layer Security}, 
      author={Thomas Knauth and Michael Steiner and Somnath Chakrabarti and Li Lei and Cedric Xing and Mona Vij},
      year={2019},
      eprint={1801.05863},
      archivePrefix={arXiv},
      primaryClass={cs.CR},
}

@misc{bhardwaj2018spx,
      title={SPX: Preserving End-to-End Security for Edge Computing}, 
      author={Ketan Bhardwaj and Ming-Wei Shih and Ada Gavrilovska and Taesoo Kim and Chengyu Song},
      year={2018},
      eprint={1809.09038},
      archivePrefix={arXiv},
      primaryClass={cs.CR},
}

@misc{intelsgx,
  title = {Intel\textsuperscript{\textcopyright} Software Guard Extensions (Intel\textsuperscript{\textcopyright} SGX)},
  howpublished = "\url{https://www.intel.com/content/www/us/en/architecture-and-technology/software-guard-extensions.html}",
}

@misc{tls,
	series =	{Request for Comments},
	number =	5246,
	howpublished =	{RFC 5246},
	publisher =	{RFC Editor},
	doi =		{10.17487/RFC5246},
	url =		{https://rfc-editor.org/rfc/rfc5246.txt},
    author =	{Eric Rescorla and Tim Dierks},
	title =		{{The Transport Layer Security (TLS) Protocol Version 1.2}},
	pagetotal =	104,
	year =		2008,
	month =		aug,
}

@misc{httpmr,
	series =	{Request for Comments},
	number =	7230,
	howpublished =	{RFC 7230},
	publisher =	{RFC Editor},
	doi =		{10.17487/RFC7230},
	url =		{https://rfc-editor.org/rfc/rfc7230.txt},
        author =	{Roy T. Fielding and Julian Reschke},
	title =		{{Hypertext Transfer Protocol (HTTP/1.1): Message Syntax and Routing}},
	pagetotal =	89,
	year =		2014,
	month =		jun,
}

@misc{httpsc,
	series =	{Request for Comments},
	number =	7231,
	howpublished =	{RFC 7231},
	publisher =	{RFC Editor},
	doi =		{10.17487/RFC7231},
	url =		{https://rfc-editor.org/rfc/rfc7231.txt},
    author =	{Roy T. Fielding and Julian Reschke},
	title =		{{Hypertext Transfer Protocol (HTTP/1.1): Semantics and Content}},
	pagetotal =	101,
	year =		2014,
	month =		jun,
}

@misc{uptls,
	series =	{Request for Comments},
	number =	2817,
	howpublished =	{RFC 2817},
	publisher =	{RFC Editor},
	doi =		{10.17487/RFC2817},
	url =		{https://rfc-editor.org/rfc/rfc2817.txt},
    author =	{Rohit Khare and Scott Lawrence},
	title =		{{Upgrading to TLS Within HTTP/1.1}},
	pagetotal =	13,
	year =		2000,
	month =		may,
}

@misc{https,
	series =	{Request for Comments},
	number =	2818,
	howpublished =	{RFC 2818},
	publisher =	{RFC Editor},
	doi =		{10.17487/RFC2818},
	url =		{https://rfc-editor.org/rfc/rfc2818.txt},
    author =	{Eric Rescorla},
	title =		{{HTTP Over TLS}},
	pagetotal =	7,
	year =		2000,
	month =		may,
}

@misc{cmp,
	series =	{Request for Comments},
	number =	4210,
	howpublished =	{RFC 4210},
	publisher =	{RFC Editor},
	doi =		{10.17487/RFC4210},
	url =		{https://rfc-editor.org/rfc/rfc4210.txt},
    author =	{Tero Mononen and Tomi Kause and Stephen Farrell and Dr. Carlisle Adams},
	title =		{{Internet X.509 Public Key Infrastructure Certificate Management Protocol (CMP)}},
	pagetotal =	95,
	year =		2005,
	month =		sep,
}

@online{preflight,
    title = "Fetch",
    url = "https://fetch.spec.whatwg.org/#cors-preflight-fetch",
}

@online{remoteattestsrv,
    title = "Remote Attestation",
    url = "https://www.intel.com/content/www/us/en/developer/tools/software-guard-extensions/attestation-services.html",
}

@online{dcapattest,
    title = "Quote Generation, Verification, and Attestation with Intel\textsuperscript{\textcopyright} Software Guard Extensions Data Center Attestation Primitives (Intel\textsuperscript{\textcopyright} SGX DCAP)",
    url = "https://www.intel.com/content/www/us/en/developer/articles/technical/quote-verification-attestation-with-intel-sgx-dcap.html",
}

@online{attestguide,
    title = "Intel\textsuperscript{\textcopyright} Software Guard Extensions ECDSA - Attestation for Data Center
Orientation Guide",
    url = "https://download.01.org/intel-sgx/dcap-1.0.1/docs/Intel_SGX_DCAP_ECDSA_Orientation.pdf",
}

\end{document}